\documentclass[english,aps,prl,preprint,showpacs,superscriptaddress]{revtex4-1}
\usepackage[T1]{fontenc}
\usepackage[latin9]{inputenc}
\usepackage{textcomp}
\usepackage{amstext}
\usepackage{amssymb}
\usepackage{graphicx}
\usepackage{esint}

\makeatletter
 
 \@ifundefined{textcolor}{}
 {%
   \definecolor{BLACK}{gray}{0}
   \definecolor{WHITE}{gray}{1}
   \definecolor{RED}{rgb}{1,0,0}
   \definecolor{GREEN}{rgb}{0,1,0}
   \definecolor{BLUE}{rgb}{0,0,1}
   \definecolor{CYAN}{cmyk}{1,0,0,0}
   \definecolor{MAGENTA}{cmyk}{0,1,0,0}
   \definecolor{YELLOW}{cmyk}{0,0,1,0}
 }

\usepackage{hyperref}
\usepackage{bm}

\let\oldhat=\hat
\renewcommand{\vec}[1]{\bm{#1}}
\renewcommand{\hat}[1]{\oldhat{\bm{#1}}}

\makeatother

\usepackage{babel}
\begin{document}

\title{Electromotive Force Generated by Spin Accumulation in FM/\texorpdfstring{$n$}{n}-GaAs
Heterostructures}

\author{C.~C.~Geppert}

\affiliation{School of Physics and Astronomy, University of Minnesota, Minneapolis,
Minnesota 55455, USA}

\email{crowell@physics.umn.edu}

\author{L.~R.~Wienkes}

\affiliation{School of Physics and Astronomy, University of Minnesota, Minneapolis,
Minnesota 55455, USA}

\author{K.~D.~Christie}

\affiliation{School of Physics and Astronomy, University of Minnesota, Minneapolis,
Minnesota 55455, USA}

\author{S.~J.~Patel}

\affiliation{Department of Electrical and Computer Engineering, University of
California, Santa Barbara, California 93106, USA}

\author{C.~J.~Palmstr\o m}

\affiliation{Department of Electrical and Computer Engineering, University of
California, Santa Barbara, California 93106, USA}

\affiliation{Department of Materials, University of California, Santa Barbara,
California 93106, USA}

\author{P.~A.~Crowell}

\affiliation{School of Physics and Astronomy, University of Minnesota, Minneapolis,
Minnesota 55455, USA}

\pacs{72.25.Dc, 72.25.Hg, 85.75.-d}
\begin{abstract}
We report on a method of quantifying spin accumulation in Co$_{2}$MnSi/$n$-GaAs
and Fe/$n$-GaAs heterostructures using a non-magnetic probe. In the
presence of a large non-equilibrium spin polarization, the combination
of a non-constant density of states and energy-dependent conductivity
generates an electromotive force (EMF). We demonstrate that this signal
dephases in the presence of applied and hyperfine fields, scales quadratically
with the polarization, and is comparable in magnitude to the spin-splitting.
Since this spin-generated EMF depends only on experimentally accessible
parameters of the bulk material, its magnitude may be used to quantify
the injected spin polarization in absolute terms.
\end{abstract}
\maketitle
Despite recent progress in demonstrating electrical spin injection
and detection in a wide variety of semiconducting materials systems\cite{Lou:2007fa,Ciorga:2009bt,Salis:2011iu,Uemura:2011db,vantErve:2007br,Appelbaum:2007ec,Dash:2009gz,Suzuki:2011in,Jansen:2012iz,Tombros:2007be,Zhou:2011ej,Saito:2013kq,Manzke:2013kq,Akiho:2013cy,Han:2013ik},
quantitative comparison among these experimental efforts is hindered
primarily by difficulties in distinguishing between bulk and interfacial
effects. All-electrical spintronic devices typically utilize ferromagnetic
(FM) elements for both the generation and detection of spin accumulation
in a non-magnetic channel. The resulting spin-dependent signal is
thus a convolution of several processes: spin-injection (interface),
spin-transport (bulk), and spin-detection (interface). Isolating the
channel or either interface for characterization requires assumptions
about the behavior of the other elements, which are not independently
measurable. One possible resolution is to detect the spin accumulation
via non-magnetic means, such as the inverse spin-Hall effect\cite{Valenzuela:2006cs,Olejnik:2012gj}.
However, to serve as the basis for comparison across several materials
systems, the parameters for any such spin-to-charge conversion must
also be well-known functions of temperature and composition.

In this letter, we report on a different method of detecting the injected
spin polarization in Co$_{2}$MnSi/$n$-GaAs and Fe/$n$-GaAs heterostructures.
In complete analogy with thermoelectric effects, the local increase
in free energy density required to establish a non-equilibrium spin
accumulation profile also generates an accompanying electromotive
force (EMF). This is the inverse of the ability of drift currents
to elongate or contract the effective spin diffusion length\cite{Yu:2002gm}.
Under open circuit conditions, this EMF may be detected as an electrostatic
potential shift by ferromagnetic and non-magnetic contacts alike.
The possibility of observing such an effect was first demonstrated
by Vera-Marun \emph{et al}. \cite{VeraMarun:2011bn,VeraMarun:2012kk}
in graphene nanostructures. We show that for degenerately doped $n$-GaAs,
this effect is greatly enhanced in the regime of large spin polarization
and may exceed the magnitude of the signals observed by traditional
ferromagnetic detection techniques. One key advantage of this spin-to-charge
conversion is that it may be characterized by independently accessible
parameters of the channel material, allowing the polarization to be
determined in absolute terms.

\begin{figure}
\includegraphics{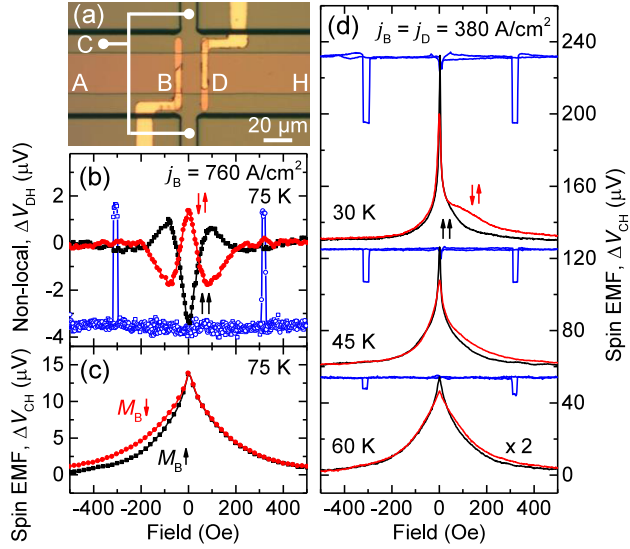}\caption{\label{fig:shortedHall}(color online). (a) Micrograph of a Co$_{2}$MnSi/$n$-GaAs
lateral spin valve device with (shorted) central Hall arms for electrostatic
detection of spin accumulation. (b) Non-local spin valve and Hanle
effect measured at contact D with forward current bias on contact
B. (c) Hanle effect observed at contact C under identical conditions
as panel (b) for both magnetization directions of contact B. (d) Spin
valve and Hanle effect data measured at contact C with identical current
biases applied simultaneously to contacts B and D.}
\end{figure}
Epitaxial (001) Co$_{2}$MnSi/$n$-GaAs and Fe/$n$-GaAs heterostructures
were grown by molecular beam epitaxy and consist of a $2.5\ \text{\textmu m}$
thick Si-doped channel $\left(n=3-5\times10^{16}\ \text{cm}^{-3}\right)$,
highly doped Schottky barrier, ($15-18\ \text{nm}$, $n^{+}=5\times10^{18}\ \text{cm}^{-3}$),
and $5\ \text{nm}$ ferromagnetic layer\cite{Lou:2007fa}. Standard
lithographic and etching techniques were used to subtractively process
lateral spin-valve devices with the pattern shown in Fig. \ref{fig:shortedHall}(a).
Two ferromagnetic contacts, labeled B and D, were positioned on either
side of a central Hall cross for the purposes of electrical spin injection
and ferromagnetic detection. Fig. \ref{fig:shortedHall}(b) shows
typical non-local spin valve and Hanle effect curves observed at contact
D for both the parallel and antiparallel configurations with a forward
current bias of $760\ \text{A}/\text{cm}^{2}$ applied to contact
B. Low-order polynomials $\left(N\le4\right)$ were subtracted from
all curves displayed in Fig. \ref{fig:shortedHall} using points outside
the regions of interest to eliminate ordinary magnetoresistive contributions.
At $75\ \text{K}$, the $d=19\ \text{\textmu m}$ center-to-center
separation between the contacts corresponds to approximately four
spin diffusion lengths $\left(\lambda_{s}=4.9\ \text{\textmu m}\right)$
as determined from standard charge and spin transport measurements
on companion devices\cite{Lou:2007fa}. The current and voltage counter-electrodes,
labeled A and H respectively, are located $240\ \text{\textmu m}$
away (not shown). These data establish the presence of a net spin
current flowing into the channel and that contact D functions properly
as a polarized detector of only one component of the non-equilibrium
spin accumulation.

This spin accumulation was also detected as a common voltage shift
in the central Hall arms, labeled C. These arms, each of length $170\ \text{\textmu m}$,
were shorted together into a single contact to eliminate any contributions
from charge- or inverse spin-Hall effects. Fig. \ref{fig:shortedHall}(c)
shows a large spin-dependent signal at contact C obtained under identical
conditions as the Hanle curves in Fig. \ref{fig:shortedHall}(b).
In contrast to the signal observed at ferromagnetic detector D, the
spin-dependent signal at contact C does not reverse sign upon reversal
of the magnetization state of the injector (contact B). A Hanle effect
of the same sign (not shown) was also observed when contact D was
used as the injector for both magnetization states. These observations
indicate that unlike the case of detection with a ferromagnet, which
is only sensitive to a particular component of the polarization, the
potential shift at contact C is sensitive to the total magnitude of
the spin accumulation in the channel. This is supported by the observation
that the width of the resulting line-shape in Fig. \ref{fig:shortedHall}(c)
matches the envelope of the Hanle curve shown in Fig. \ref{fig:shortedHall}(a).

To further demonstrate that the observed potential shift depends on
the total magnitude of the spin accumulation in the channel (although
not its overall sign), two separate $380\ \text{A}/\text{cm}^{2}$
forward current biases were applied simultaneously to contacts B and
D with contact A serving as the counter-electrode for both current
sources. With both ferromagnetic contacts functioning as spin injectors,
the total magnitude of the spin accumulation in the channel was increased
(decreased) by aligning the magnetization of the two contacts parallel
(anti-parallel) to each other\cite{VeraMarun:2012kk}. As shown in
Fig. \ref{fig:shortedHall}(d), this allows observation of a spin
valve effect at non-magnetic probe C when a magnetic field is swept
along the in-plane ferromagnetic easy axis {[}110{]}. The switching
events occurring at $\approx300\ \text{Oe}$ are the same as those
observed in Fig. \ref{fig:shortedHall}(b).

Fig. \ref{fig:shortedHall}(d) shows the corresponding Hanle curves
obtained by sweeping the magnetic field normal to the plane of the
device. The difference between the parallel and anti-parallel Hanle
curves at $0\ \text{Oe}$ equals that of the spin valve signal, demonstrating
that the signal at contact C is sensitive to the superposition of
the spin accumulation arising from both ferromagnetic injectors B
and D. The Hanle line-shape broadens at higher temperatures as expected
due to an increased spin relaxation rate, with both Hanle curves trending
to zero at higher applied field.

\begin{figure}
\includegraphics{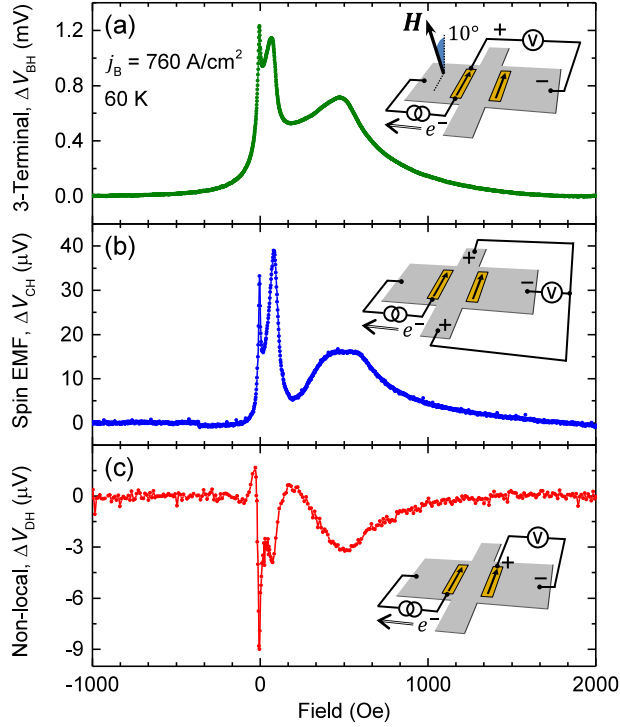}\caption{\label{fig:obliqueHanle}(color online). Oblique Hanle effects observed
at (a) ferromagnetic injector B (three-terminal), (b) semiconducting
arms C (spin-generated EMF), and (c) remote ferromagnetic detector
D (non-local) under identical conditions with forward current bias
on contact B.}
\end{figure}
The additional low-field features in Fig. \ref{fig:shortedHall}(d)
at low temperatures are caused by hyperfine interactions with dynamically
polarized nuclei, which exert large effective Overhauser fields on
the electron spin dynamics\cite{Paget:1981td,Kolbl:2012gi}. This
necessitates a more complex interpretation of Hanle line-shapes\cite{Chan:2009bu},
yet also makes possible a robust test of whether a particular signal
originates from spin-dependent processes. In the oblique Hanle geometry
shown in Fig. \ref{fig:obliqueHanle}, the magnetic field is applied
at an intermediate angle between the device normal and the ferromagnetic
easy axis. In this configuration, two additional satellite peaks appear
at field values where the Overhauser field partially or wholly cancels
the applied field. This cancellation reduces the Hanle dephasing effect
and allows the electron spin ensemble to re-polarize. Fig. \ref{fig:obliqueHanle}(a),
(b) and (c) show the voltage measured on a second Co$_{2}$MnSi device
at contacts B (three-terminal), C (spin-generated EMF), and D (non-local)
respectively as the field is swept at a constant $10$\textdegree{}
orientation from the device normal. The appearance of satellite peaks
at comparable positions in all three measurements provides conclusive
evidence that the voltage shift observed at contact C is a direct
measure of the spin accumulation in the channel. Since the sign of
the Overhauser field in bulk GaAs is known\cite{Paget:1977ge}, the
position of the satellite peaks at positive field indicates majority
spin accumulation in the channel. Note that the sign of the Hanle
curve in Fig. \ref{fig:obliqueHanle}(a) is opposite that of Fig.\ref{fig:obliqueHanle}(c)
due to the inverted sign of the ferromagnetic detection efficiency
at high forward bias\cite{Crooker:2009ju}.

\begin{figure}
\includegraphics{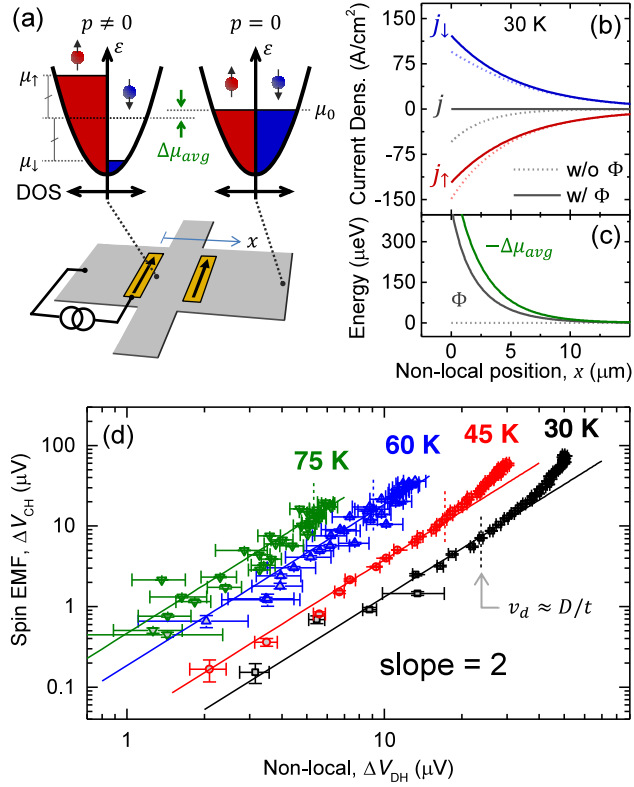}\caption{\label{fig:powerLaw}(color online). (a) Diagram of chemical potentials
in the presence (left) and absence (right) of spin accumulation. (b)
Current density as a function of non-local position for each spin
band ($\uparrow$ red, $\downarrow$ blue) and their total (black).
Dotted (solid) curves exclude (include) the contribution from the
steady-state potential $\Phi$. (c) Magnitude of $\Phi$ and average
chemical potential shift $\Delta\mu_{avg}$ vs. non-local position.
(d) Log-log plot of the spin-generated EMF observed at contact C vs.
(ferromagnetic) non-local spin-valve magnitude at contact D showing
a quadratic dependence at low biases. Solid lines have a slope of
2.}
\end{figure}
The current density for spin-up (spin-down) electrons is given by
the gradient of the total electrochemical potential as $e\vec{j}_{\uparrow\left(\downarrow\right)}=\sigma_{\uparrow\left(\downarrow\right)}\vec{\nabla}\left(\mu_{\uparrow\left(\downarrow\right)}-e\Phi\right)$.
$\mu_{\uparrow\left(\downarrow\right)}$ denotes the number density-dependent
chemical potentials and $\Phi$ is the electrostatic potential common
to both spin-bands. Following the approach in Ref. \cite{Yu:2002gm},
we make the assumption that the conductivity of each band $\sigma_{\uparrow\left(\downarrow\right)}=n_{\uparrow\left(\downarrow\right)}e\nu$
is proportional to the number density $n_{\uparrow\left(\downarrow\right)}$
with a spin-independent mobility $\nu$. This allows the net charge
current density $\vec{j}=\vec{j}_{\uparrow}+\vec{j}_{\downarrow}$
to be written as
\begin{equation}
\frac{\vec{j}}{\sigma}=\vec{\nabla}\left(\frac{\mu_{\uparrow}+\mu_{\downarrow}}{2e}\right)+p\vec{\nabla}\left(\frac{\mu_{\uparrow}-\mu_{\downarrow}}{2e}\right)-\vec{\nabla}\Phi,\label{eq:currentDensity}
\end{equation}
where $p=\left(n_{\uparrow}-n_{\downarrow}\right)/n$ is the fractional
number polarization, $n$ is the total carrier concentration, and
$\sigma=ne\nu$ is the channel conductivity. Due to the presence of
the first two terms, a `pure' non-local spin current ($\vec{j}_{\uparrow}=-\vec{j}_{\downarrow}$)
cannot be achieved without establishing a non-zero electrostatic potential
gradient $\left(\vec{\nabla}\Phi\ne0\right)$. In the presence of
a spin accumulation, the first term of Eq. \ref{eq:currentDensity}
is non-zero due to the asymmetric shift of $\mu_{\uparrow}$ and $\mu_{\downarrow}$
relative to the unpolarized state. This is shown schematically in
Fig. \ref{fig:powerLaw}(a), where the decrease in $\Delta\mu_{avg}=\left(\mu_{\uparrow}+\mu_{\downarrow}\right)/2-\mu_{0}$
is due to the non-constant density of states. The second term in Eq.
\ref{eq:currentDensity} originates from the energy dependence of
the conductivity, which for the Drude form above may be expressed
in terms of the population imbalance of the two spin sub-bands. The
dashed lines in Fig. \ref{fig:powerLaw}(b) show the asymmetric nature
of $j_{\uparrow}$ and $j_{\downarrow}$ for a typical polarization
profile $p=p\left(0\right)e^{-x/\lambda_{s}}$ at 30 K ($p\left(0\right)=0.6$,
$\lambda_{s}=5.6\ \text{nm}$) in the absence of an electrostatic
potential gradient $\left(\vec{\nabla}\Phi=0\right)$. Here $x$ denotes
the (non-local) position to the right of the injector contact. The
solid lines indicate the steady-state condition after taking into
account the spin-generated EMF.

For small polarizations $\left(p\ll1\right)$, we can Taylor expand
$\mu_{\uparrow}=\mu\left(2n_{\uparrow}\right)$ and $\mu_{\downarrow}=\mu\left(2n_{\downarrow}\right)$
about the background concentration $n$:
\begin{equation}
\mu_{\uparrow\left(\downarrow\right)}\approx\mu_{0}+\left(-\right)\frac{\partial\mu}{\partial n}np+\frac{1}{2}\frac{\partial^{2}\mu}{\partial n^{2}}n^{2}p^{2},\label{eq:taylorExpand}
\end{equation}
where $2n_{\uparrow\left(\downarrow\right)}=\left[1+\left(-\right)p\right]n$.
This allows Eq. \ref{eq:currentDensity} to be simplified as
\begin{equation}
\vec{j}=\sigma\vec{\nabla}\left(kp^{2}-\Phi\right),\label{eq:kp2}
\end{equation}
with 
\begin{equation}
k=\frac{1}{2e}\left(\frac{\partial^{2}\mu}{\partial n^{2}}n^{2}+\frac{\partial\mu}{\partial n}n\right)=\frac{n}{2}\frac{\partial}{\partial n}\left(\frac{D}{\nu}\right),\label{eq:kSimple}
\end{equation}
where the diffusion constant is defined via the Einstein relation
$eD=n\nu\left(\partial\mu/\partial n\right)$ \cite{Ashcroft:1976ud}.
The first term in Eq. \ref{eq:kSimple} is analogous to the first
order term in the Sommerfeld expansion used to analyze the Seebeck
effect\cite{Ashcroft:1976ud,Johnson:1953ud}. The second term, absent
in the usual Seebeck analysis, appears as a consequence of the imbalance
in the number of carriers in the two spin sub-bands and is formally
equivalent to the contribution discussed in Ref. \cite{VeraMarun:2011bn}.
For a parabolic density of states with effective mass $m^{\star}$,
the function $\mu\left(n\right)$ may be obtained by numerically inverting
the usual relation $n=n_{Q}\mathcal{F}_{1/2}\left[\mu/k_{B}T\right]$,
where $n_{Q}=2\left(m^{\star}k_{B}T/2\hbar^{2}\pi\right)^{3/2}$ is
the quantum concentration and $\mathcal{F}_{1/2}\left(\xi\right)=\left(2/\sqrt{\pi}\right)\int_{0}^{\infty}x^{1/2}\left[e^{x-\xi}+1\right]^{-1}dx$
is the Fermi-Dirac integral\cite{Smith:1978wp}. The pre-factor $k$
takes on a value of $\frac{2}{9}\varepsilon_{F}/e$ in the degenerate
case $\left(n\gg n_{Q}\right)$ and vanishes in the non-degenerate
limit $\left(n\ll n_{Q}\right)$. Typical values for $\Phi$ and $\Delta\mu_{avg}$
are shown in Fig. \ref{fig:powerLaw}(c) for a concentration of $n=3\times10^{16}\ \text{cm}^{-3}$
(Fermi energy, $\varepsilon_{F}=5.3\ \text{meV}$) at $30\ \text{K}$.

In the non-local region where $j=0$, Eq. \ref{eq:kp2} gives the
electrostatic potential as $\Phi=kp^{2}$. This may be regarded as
a source of EMF which scales \emph{quadratically} with the spin accumulation.
This is in contrast to the non-local spin valve signal which scales
\emph{linearly}: $e\Delta V_{DH}=\eta_{0}\left(\mu_{\uparrow}-\mu_{\downarrow}\right)\approx2\eta_{0}\frac{\partial\mu}{\partial n}p$,
where $\eta_{0}$ is the FM spin detection efficiency. Fig. \ref{fig:powerLaw}(d)
shows the Hanle and spin valve magnitudes observed at contacts C and
D plotted against each other on a log-log plot while varying the current
bias applied to injector contact B. The data at low bias confirm the
predicted slope of two. The departure from quadratic behavior at high
bias is caused by drift currents which redistribute more of the polarization
toward the interfacial region which, as discussed below, sets the
boundary condition for observing the spin-generated EMF at contact
C. The onset of this drift effect occurs when the effective spin diffusion
length\cite{Yu:2002gm} becomes smaller than the channel thickness
$t$, i.e. drift velocity $v_{d}=j/ne\approx D/t$ as indicated by
the vertical dashed lines in Fig. \ref{fig:powerLaw}(d).

\begin{figure}
\includegraphics{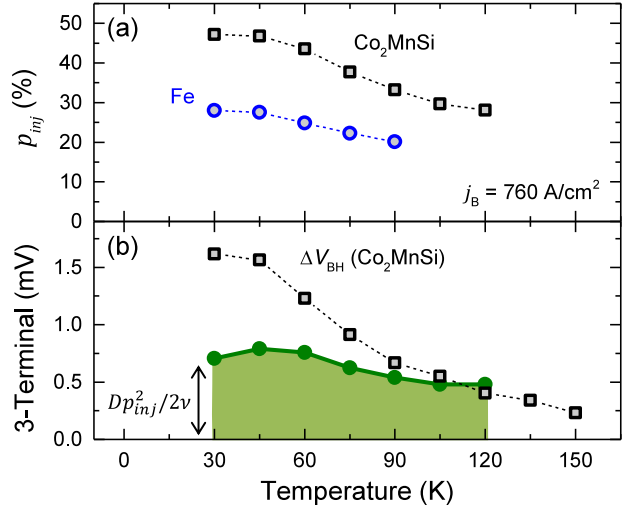}\caption{\label{fig:polarizationVsTemp}(color online). (a) Polarization vs.
temperature at injector contact B determined from the magnitude of
the ES shift (Co$_{2}$MnSi: black squares, Fe: blue circles). (b)
Temperature dependence of experimental three-terminal signal (Co$_{2}$MnSi:
black squares). The green shaded region shows the average electrochemical
shift at the injector determined from the Co$_{2}$MnSi polarization
values in panel (a).}
\end{figure}
To extract an absolute polarization from the magnitude of the voltage
shift observed at contact C, we note that the electric field within
the metallic interfacial regions is zero $\left(\hat{n}\times\vec{\nabla}\Phi=\vec{0}\right)$.
Here $\hat{n}$ is a unit vector normal to the surfaces of the contact
regions B and D. The importance of the this condition is two-fold.
First, it allows the shift in $\Phi$ to be non-zero even in regions
where $p=0$. The $n^{+}$ layers beneath contacts B and D effectively
short out any transverse potential drop\cite{Isenberg:1948jk} across
the width of the channel, allowing observation of a signal at contact
C. Second, because the shorting effect causes the shift in $\Phi$
to differ from $kp^{2}$, Eq. \ref{eq:kp2} implies the existence
of divergenceless eddy currents sustained by the non-equilibrium spin
polarization. These eddy currents circulate through the $n^{+}$ region
into the bulk without significantly disturbing the spin accumulation
profile since $kp^{2}\ll D/\nu$.

We determine the spin accumulation profile in the channel by discretizing
the standard spin drift-diffusion relations $\left(\vec{v}_{d}\cdot\vec{\nabla}p+D\nabla^{2}p+Dp/\lambda_{s}^{2}=0\right)$
in three dimensions and solving for the average injector polarization
$p_{inj}$ which upon Laplace relaxation yields the experimentally
observed EMF%
\footnote{See Supplemental Material at {[}URL from Publisher{]} for calculation
details.%
}. Fig. \ref{fig:polarizationVsTemp}(a) shows $p_{inj}$ as a function
of temperature for both a Co$_{2}$MnSi device ($3\times10^{16}\ \text{cm}^{-3}$)
and an Fe device ($5\times10^{16}\ \text{cm}^{-3}$). Notably, these
values were obtained solely from bulk semiconductor parameters and
are independent of any assumptions about the efficiency of spin-dependent
tunneling at the heterojuctions.

The experimental temperature dependence of the three-terminal signal
is shown in Fig. \ref{fig:polarizationVsTemp}(b), where injector
contact B also functions as a polarized detector. The magnitude of
the three-terminal Hanle signal may be decomposed as $\Delta V_{\text{BH}}\approx\left(D/\nu\right)\left(\eta p_{inj}+p_{inj}^{2}/2\right)$.
The first term originates from the usual spin-dependence of the interfacial
tunneling conductance. The second term represents a hitherto unconsidered
contribution from the spin-generated EMF as a shift of the average
electrochemical potential. The green shaded region in Fig. \ref{fig:polarizationVsTemp}(b)
indicates the magnitude of this contribution as determined from the
polarization values in Fig. \ref{fig:polarizationVsTemp}(a). The
spin-generated EMF constitutes a significant fraction of the three-terminal
signal across the entire temperature range. This contribution must
therefore be taken into account when interpreting ferromagnetic detection
signals in the regime of large spin polarization.

In conclusion, we have experimentally demonstrated the presence of
an EMF which must accompany a spin accumulation in steady-state due
to a non-constant density of states and population imbalance of the
two spin sub-bands. The close analogy with thermoelectric physics
suggests the possibility of observing a class of related effects in
which the source of the deviation from equilibrium in free energy
density is spin rather than heat. Since the behavior of this phenomenon
may be parameterized by independent measurements, the spin accumulation
in the channel may be identified unambiguously and quantified in absolute
terms. This represents an attractive detection alternative in situations
where traditional ferromagnetic techniques are either unfeasible or
unreliable.

This work was supported by NSF Grant No. DMR-1104951; by C-SPIN, one
of the six centers of STARnet, a Semiconductor Research Corporation
program sponsored by MARCO and DARPA; and by the NSF MRSEC program.
Parts of this work were carried out in the Minnesota Nano Center which
receives partial support from NSF through the NNIN program.\bibliographystyle{apsrev4-1}
\bibliography{p2paper}

\begin{thebibliography}{30}%
\makeatletter
\providecommand \@ifxundefined [1]{%
 \@ifx{#1\undefined}
}%
\providecommand \@ifnum [1]{%
 \ifnum #1\expandafter \@firstoftwo
 \else \expandafter \@secondoftwo
 \fi
}%
\providecommand \@ifx [1]{%
 \ifx #1\expandafter \@firstoftwo
 \else \expandafter \@secondoftwo
 \fi
}%
\providecommand \natexlab [1]{#1}%
\providecommand \enquote  [1]{``#1''}%
\providecommand \bibnamefont  [1]{#1}%
\providecommand \bibfnamefont [1]{#1}%
\providecommand \citenamefont [1]{#1}%
\providecommand \href@noop [0]{\@secondoftwo}%
\providecommand \href [0]{\begingroup \@sanitize@url \@href}%
\providecommand \@href[1]{\@@startlink{#1}\@@href}%
\providecommand \@@href[1]{\endgroup#1\@@endlink}%
\providecommand \@sanitize@url [0]{\catcode `\\12\catcode `\$12\catcode
  `\&12\catcode `\#12\catcode `\^12\catcode `\_12\catcode `\%12\relax}%
\providecommand \@@startlink[1]{}%
\providecommand \@@endlink[0]{}%
\providecommand \url  [0]{\begingroup\@sanitize@url \@url }%
\providecommand \@url [1]{\endgroup\@href {#1}{\urlprefix }}%
\providecommand \urlprefix  [0]{URL }%
\providecommand \Eprint [0]{\href }%
\providecommand \doibase [0]{http://dx.doi.org/}%
\providecommand \selectlanguage [0]{\@gobble}%
\providecommand \bibinfo  [0]{\@secondoftwo}%
\providecommand \bibfield  [0]{\@secondoftwo}%
\providecommand \translation [1]{[#1]}%
\providecommand \BibitemOpen [0]{}%
\providecommand \bibitemStop [0]{}%
\providecommand \bibitemNoStop [0]{.\EOS\space}%
\providecommand \EOS [0]{\spacefactor3000\relax}%
\providecommand \BibitemShut  [1]{\csname bibitem#1\endcsname}%
\let\auto@bib@innerbib\@empty
\bibitem [{\citenamefont {Lou}\ \emph {et~al.}(2007)\citenamefont {Lou},
  \citenamefont {Adelmann}, \citenamefont {Crooker}, \citenamefont {Garlid},
  \citenamefont {Zhang}, \citenamefont {Reddy}, \citenamefont {Flexner},
  \citenamefont {Palmstr{\o}m},\ and\ \citenamefont {Crowell}}]{Lou:2007fa}%
  \BibitemOpen
  \bibfield  {author} {\bibinfo {author} {\bibfnamefont {X.}~\bibnamefont
  {Lou}}, \bibinfo {author} {\bibfnamefont {C.}~\bibnamefont {Adelmann}},
  \bibinfo {author} {\bibfnamefont {S.~A.}\ \bibnamefont {Crooker}}, \bibinfo
  {author} {\bibfnamefont {E.~S.}\ \bibnamefont {Garlid}}, \bibinfo {author}
  {\bibfnamefont {J.}~\bibnamefont {Zhang}}, \bibinfo {author} {\bibfnamefont
  {K.~S.~M.}\ \bibnamefont {Reddy}}, \bibinfo {author} {\bibfnamefont {S.~D.}\
  \bibnamefont {Flexner}}, \bibinfo {author} {\bibfnamefont {C.~J.}\
  \bibnamefont {Palmstr{\o}m}}, \ and\ \bibinfo {author} {\bibfnamefont
  {P.~A.}\ \bibnamefont {Crowell}},\ }\href@noop {} {\bibfield  {journal}
  {\bibinfo  {journal} {Nature Phys.}\ }\textbf {\bibinfo {volume} {3}},\
  \bibinfo {pages} {197} (\bibinfo {year} {2007})}\BibitemShut {NoStop}%
\bibitem [{\citenamefont {Ciorga}\ \emph {et~al.}(2009)\citenamefont {Ciorga},
  \citenamefont {Einwanger}, \citenamefont {Wurstbauer}, \citenamefont {Schuh},
  \citenamefont {Wegscheider},\ and\ \citenamefont {Weiss}}]{Ciorga:2009bt}%
  \BibitemOpen
  \bibfield  {author} {\bibinfo {author} {\bibfnamefont {M.}~\bibnamefont
  {Ciorga}}, \bibinfo {author} {\bibfnamefont {A.}~\bibnamefont {Einwanger}},
  \bibinfo {author} {\bibfnamefont {U.}~\bibnamefont {Wurstbauer}}, \bibinfo
  {author} {\bibfnamefont {D.}~\bibnamefont {Schuh}}, \bibinfo {author}
  {\bibfnamefont {W.}~\bibnamefont {Wegscheider}}, \ and\ \bibinfo {author}
  {\bibfnamefont {D.}~\bibnamefont {Weiss}},\ }\href@noop {} {\bibfield
  {journal} {\bibinfo  {journal} {Phys. Rev. B}\ }\textbf {\bibinfo {volume}
  {79}},\ \bibinfo {pages} {165321} (\bibinfo {year} {2009})}\BibitemShut
  {NoStop}%
\bibitem [{\citenamefont {Salis}\ \emph {et~al.}(2011)\citenamefont {Salis},
  \citenamefont {Alvarado},\ and\ \citenamefont {Fuhrer}}]{Salis:2011iu}%
  \BibitemOpen
  \bibfield  {author} {\bibinfo {author} {\bibfnamefont {G.}~\bibnamefont
  {Salis}}, \bibinfo {author} {\bibfnamefont {S.~F.}\ \bibnamefont {Alvarado}},
  \ and\ \bibinfo {author} {\bibfnamefont {A.}~\bibnamefont {Fuhrer}},\
  }\href@noop {} {\bibfield  {journal} {\bibinfo  {journal} {Phys. Rev. B}\
  }\textbf {\bibinfo {volume} {84}},\ \bibinfo {pages} {041307} (\bibinfo
  {year} {2011})}\BibitemShut {NoStop}%
\bibitem [{\citenamefont {Uemura}\ \emph {et~al.}(2011)\citenamefont {Uemura},
  \citenamefont {Akiho}, \citenamefont {Harada}, \citenamefont {Matsuda},\ and\
  \citenamefont {Yamamoto}}]{Uemura:2011db}%
  \BibitemOpen
  \bibfield  {author} {\bibinfo {author} {\bibfnamefont {T.}~\bibnamefont
  {Uemura}}, \bibinfo {author} {\bibfnamefont {T.}~\bibnamefont {Akiho}},
  \bibinfo {author} {\bibfnamefont {M.}~\bibnamefont {Harada}}, \bibinfo
  {author} {\bibfnamefont {K.-i.}\ \bibnamefont {Matsuda}}, \ and\ \bibinfo
  {author} {\bibfnamefont {M.}~\bibnamefont {Yamamoto}},\ }\href@noop {}
  {\bibfield  {journal} {\bibinfo  {journal} {Appl. Phys. Lett.}\ }\textbf
  {\bibinfo {volume} {99}},\ \bibinfo {pages} {082108} (\bibinfo {year}
  {2011})}\BibitemShut {NoStop}%
\bibitem [{\citenamefont {van~'t Erve}\ \emph {et~al.}(2007)\citenamefont
  {van~'t Erve}, \citenamefont {Hanbicki}, \citenamefont {Holub}, \citenamefont
  {Li}, \citenamefont {Awo-Affouda}, \citenamefont {Thompson},\ and\
  \citenamefont {Jonker}}]{vantErve:2007br}%
  \BibitemOpen
  \bibfield  {author} {\bibinfo {author} {\bibfnamefont {O.~M.~J.}\
  \bibnamefont {van~'t Erve}}, \bibinfo {author} {\bibfnamefont {A.~T.}\
  \bibnamefont {Hanbicki}}, \bibinfo {author} {\bibfnamefont {M.}~\bibnamefont
  {Holub}}, \bibinfo {author} {\bibfnamefont {C.~H.}\ \bibnamefont {Li}},
  \bibinfo {author} {\bibfnamefont {C.}~\bibnamefont {Awo-Affouda}}, \bibinfo
  {author} {\bibfnamefont {P.~E.}\ \bibnamefont {Thompson}}, \ and\ \bibinfo
  {author} {\bibfnamefont {B.~T.}\ \bibnamefont {Jonker}},\ }\href@noop {}
  {\bibfield  {journal} {\bibinfo  {journal} {Appl. Phys. Lett.}\ }\textbf
  {\bibinfo {volume} {91}},\ \bibinfo {pages} {212109} (\bibinfo {year}
  {2007})}\BibitemShut {NoStop}%
\bibitem [{\citenamefont {Appelbaum}\ \emph {et~al.}(2007)\citenamefont
  {Appelbaum}, \citenamefont {Huang},\ and\ \citenamefont
  {Monsma}}]{Appelbaum:2007ec}%
  \BibitemOpen
  \bibfield  {author} {\bibinfo {author} {\bibfnamefont {I.}~\bibnamefont
  {Appelbaum}}, \bibinfo {author} {\bibfnamefont {B.}~\bibnamefont {Huang}}, \
  and\ \bibinfo {author} {\bibfnamefont {D.~J.}\ \bibnamefont {Monsma}},\
  }\href@noop {} {\bibfield  {journal} {\bibinfo  {journal} {Nature}\ }\textbf
  {\bibinfo {volume} {447}},\ \bibinfo {pages} {295} (\bibinfo {year}
  {2007})}\BibitemShut {NoStop}%
\bibitem [{\citenamefont {Dash}\ \emph {et~al.}(2009)\citenamefont {Dash},
  \citenamefont {Sharma}, \citenamefont {Patel}, \citenamefont {de~Jong},\ and\
  \citenamefont {Jansen}}]{Dash:2009gz}%
  \BibitemOpen
  \bibfield  {author} {\bibinfo {author} {\bibfnamefont {S.~P.}\ \bibnamefont
  {Dash}}, \bibinfo {author} {\bibfnamefont {S.}~\bibnamefont {Sharma}},
  \bibinfo {author} {\bibfnamefont {R.~S.}\ \bibnamefont {Patel}}, \bibinfo
  {author} {\bibfnamefont {M.~P.}\ \bibnamefont {de~Jong}}, \ and\ \bibinfo
  {author} {\bibfnamefont {R.}~\bibnamefont {Jansen}},\ }\href@noop {}
  {\bibfield  {journal} {\bibinfo  {journal} {Nature}\ }\textbf {\bibinfo
  {volume} {462}},\ \bibinfo {pages} {491} (\bibinfo {year}
  {2009})}\BibitemShut {NoStop}%
\bibitem [{\citenamefont {Suzuki}\ \emph {et~al.}(2011)\citenamefont {Suzuki},
  \citenamefont {Sasaki}, \citenamefont {Oikawa}, \citenamefont {Shiraishi},
  \citenamefont {Suzuki},\ and\ \citenamefont {Noguchi}}]{Suzuki:2011in}%
  \BibitemOpen
  \bibfield  {author} {\bibinfo {author} {\bibfnamefont {T.}~\bibnamefont
  {Suzuki}}, \bibinfo {author} {\bibfnamefont {T.}~\bibnamefont {Sasaki}},
  \bibinfo {author} {\bibfnamefont {T.}~\bibnamefont {Oikawa}}, \bibinfo
  {author} {\bibfnamefont {M.}~\bibnamefont {Shiraishi}}, \bibinfo {author}
  {\bibfnamefont {Y.}~\bibnamefont {Suzuki}}, \ and\ \bibinfo {author}
  {\bibfnamefont {K.}~\bibnamefont {Noguchi}},\ }\href@noop {} {\bibfield
  {journal} {\bibinfo  {journal} {Appl. Phys. Express}\ }\textbf {\bibinfo
  {volume} {4}},\ \bibinfo {pages} {023003} (\bibinfo {year}
  {2011})}\BibitemShut {NoStop}%
\bibitem [{\citenamefont {Jansen}\ \emph {et~al.}(2012)\citenamefont {Jansen},
  \citenamefont {Dash}, \citenamefont {Sharma},\ and\ \citenamefont
  {Min}}]{Jansen:2012iz}%
  \BibitemOpen
  \bibfield  {author} {\bibinfo {author} {\bibfnamefont {R.}~\bibnamefont
  {Jansen}}, \bibinfo {author} {\bibfnamefont {S.~P.}\ \bibnamefont {Dash}},
  \bibinfo {author} {\bibfnamefont {S.}~\bibnamefont {Sharma}}, \ and\ \bibinfo
  {author} {\bibfnamefont {B.~C.}\ \bibnamefont {Min}},\ }\href@noop {}
  {\bibfield  {journal} {\bibinfo  {journal} {Semicond. Sci. Technol.}\
  }\textbf {\bibinfo {volume} {27}},\ \bibinfo {pages} {083001} (\bibinfo
  {year} {2012})}\BibitemShut {NoStop}%
\bibitem [{\citenamefont {Tombros}\ \emph {et~al.}(2007)\citenamefont
  {Tombros}, \citenamefont {Jozsa}, \citenamefont {Popinciuc}, \citenamefont
  {Jonkman},\ and\ \citenamefont {van Wees}}]{Tombros:2007be}%
  \BibitemOpen
  \bibfield  {author} {\bibinfo {author} {\bibfnamefont {N.}~\bibnamefont
  {Tombros}}, \bibinfo {author} {\bibfnamefont {C.}~\bibnamefont {Jozsa}},
  \bibinfo {author} {\bibfnamefont {M.}~\bibnamefont {Popinciuc}}, \bibinfo
  {author} {\bibfnamefont {H.~T.}\ \bibnamefont {Jonkman}}, \ and\ \bibinfo
  {author} {\bibfnamefont {B.~J.}\ \bibnamefont {van Wees}},\ }\href@noop {}
  {\bibfield  {journal} {\bibinfo  {journal} {Nature}\ }\textbf {\bibinfo
  {volume} {448}},\ \bibinfo {pages} {571} (\bibinfo {year}
  {2007})}\BibitemShut {NoStop}%
\bibitem [{\citenamefont {Zhou}\ \emph {et~al.}(2011)\citenamefont {Zhou},
  \citenamefont {Han}, \citenamefont {Chang}, \citenamefont {Xiu},
  \citenamefont {Wang}, \citenamefont {Oehme}, \citenamefont {Fischer},
  \citenamefont {Schulze}, \citenamefont {Kawakami},\ and\ \citenamefont
  {Wang}}]{Zhou:2011ej}%
  \BibitemOpen
  \bibfield  {author} {\bibinfo {author} {\bibfnamefont {Y.}~\bibnamefont
  {Zhou}}, \bibinfo {author} {\bibfnamefont {W.}~\bibnamefont {Han}}, \bibinfo
  {author} {\bibfnamefont {L.-T.}\ \bibnamefont {Chang}}, \bibinfo {author}
  {\bibfnamefont {F.}~\bibnamefont {Xiu}}, \bibinfo {author} {\bibfnamefont
  {M.}~\bibnamefont {Wang}}, \bibinfo {author} {\bibfnamefont {M.}~\bibnamefont
  {Oehme}}, \bibinfo {author} {\bibfnamefont {I.~A.}\ \bibnamefont {Fischer}},
  \bibinfo {author} {\bibfnamefont {J.}~\bibnamefont {Schulze}}, \bibinfo
  {author} {\bibfnamefont {R.~K.}\ \bibnamefont {Kawakami}}, \ and\ \bibinfo
  {author} {\bibfnamefont {K.~L.}\ \bibnamefont {Wang}},\ }\href@noop {}
  {\bibfield  {journal} {\bibinfo  {journal} {Phys. Rev. B}\ }\textbf {\bibinfo
  {volume} {84}},\ \bibinfo {pages} {125323} (\bibinfo {year}
  {2011})}\BibitemShut {NoStop}%
\bibitem [{\citenamefont {Saito}\ \emph {et~al.}(2013)\citenamefont {Saito},
  \citenamefont {Tezuka}, \citenamefont {Matsuura},\ and\ \citenamefont
  {Sugimoto}}]{Saito:2013kq}%
  \BibitemOpen
  \bibfield  {author} {\bibinfo {author} {\bibfnamefont {T.}~\bibnamefont
  {Saito}}, \bibinfo {author} {\bibfnamefont {N.}~\bibnamefont {Tezuka}},
  \bibinfo {author} {\bibfnamefont {M.}~\bibnamefont {Matsuura}}, \ and\
  \bibinfo {author} {\bibfnamefont {S.}~\bibnamefont {Sugimoto}},\ }\href@noop
  {} {\bibfield  {journal} {\bibinfo  {journal} {Appl. Phys. Lett.}\ }\textbf
  {\bibinfo {volume} {103}},\ \bibinfo {pages} {122401} (\bibinfo {year}
  {2013})}\BibitemShut {NoStop}%
\bibitem [{\citenamefont {Manzke}\ \emph {et~al.}(2013)\citenamefont {Manzke},
  \citenamefont {Farshchi}, \citenamefont {Bruski}, \citenamefont {Herfort},\
  and\ \citenamefont {Ramsteiner}}]{Manzke:2013kq}%
  \BibitemOpen
  \bibfield  {author} {\bibinfo {author} {\bibfnamefont {Y.}~\bibnamefont
  {Manzke}}, \bibinfo {author} {\bibfnamefont {R.}~\bibnamefont {Farshchi}},
  \bibinfo {author} {\bibfnamefont {P.}~\bibnamefont {Bruski}}, \bibinfo
  {author} {\bibfnamefont {J.}~\bibnamefont {Herfort}}, \ and\ \bibinfo
  {author} {\bibfnamefont {M.}~\bibnamefont {Ramsteiner}},\ }\href@noop {}
  {\bibfield  {journal} {\bibinfo  {journal} {Phys. Rev. B}\ }\textbf {\bibinfo
  {volume} {87}},\ \bibinfo {pages} {134415} (\bibinfo {year}
  {2013})}\BibitemShut {NoStop}%
\bibitem [{\citenamefont {Akiho}\ \emph {et~al.}(2013)\citenamefont {Akiho},
  \citenamefont {Shan}, \citenamefont {Liu}, \citenamefont {Matsuda},
  \citenamefont {Yamamoto},\ and\ \citenamefont {Uemura}}]{Akiho:2013cy}%
  \BibitemOpen
  \bibfield  {author} {\bibinfo {author} {\bibfnamefont {T.}~\bibnamefont
  {Akiho}}, \bibinfo {author} {\bibfnamefont {J.}~\bibnamefont {Shan}},
  \bibinfo {author} {\bibfnamefont {H.-x.}\ \bibnamefont {Liu}}, \bibinfo
  {author} {\bibfnamefont {K.-i.}\ \bibnamefont {Matsuda}}, \bibinfo {author}
  {\bibfnamefont {M.}~\bibnamefont {Yamamoto}}, \ and\ \bibinfo {author}
  {\bibfnamefont {T.}~\bibnamefont {Uemura}},\ }\href@noop {} {\bibfield
  {journal} {\bibinfo  {journal} {Phys. Rev. B}\ }\textbf {\bibinfo {volume}
  {87}},\ \bibinfo {pages} {235205} (\bibinfo {year} {2013})}\BibitemShut
  {NoStop}%
\bibitem [{\citenamefont {Han}\ \emph {et~al.}(2013)\citenamefont {Han},
  \citenamefont {Jiang}, \citenamefont {Kajdos}, \citenamefont {Yang},
  \citenamefont {Stemmer},\ and\ \citenamefont {Parkin}}]{Han:2013ik}%
  \BibitemOpen
  \bibfield  {author} {\bibinfo {author} {\bibfnamefont {W.}~\bibnamefont
  {Han}}, \bibinfo {author} {\bibfnamefont {X.}~\bibnamefont {Jiang}}, \bibinfo
  {author} {\bibfnamefont {A.}~\bibnamefont {Kajdos}}, \bibinfo {author}
  {\bibfnamefont {S.-H.}\ \bibnamefont {Yang}}, \bibinfo {author}
  {\bibfnamefont {S.}~\bibnamefont {Stemmer}}, \ and\ \bibinfo {author}
  {\bibfnamefont {S.~S.~P.}\ \bibnamefont {Parkin}},\ }\href@noop {} {\bibfield
   {journal} {\bibinfo  {journal} {Nature Communications}\ }\textbf {\bibinfo
  {volume} {4}},\ \bibinfo {pages} {1} (\bibinfo {year} {2013})}\BibitemShut
  {NoStop}%
\bibitem [{\citenamefont {Valenzuela}\ and\ \citenamefont
  {Tinkham}(2006)}]{Valenzuela:2006cs}%
  \BibitemOpen
  \bibfield  {author} {\bibinfo {author} {\bibfnamefont {S.~O.}\ \bibnamefont
  {Valenzuela}}\ and\ \bibinfo {author} {\bibfnamefont {M.}~\bibnamefont
  {Tinkham}},\ }\href@noop {} {\bibfield  {journal} {\bibinfo  {journal}
  {Nature}\ }\textbf {\bibinfo {volume} {442}},\ \bibinfo {pages} {176}
  (\bibinfo {year} {2006})}\BibitemShut {NoStop}%
\bibitem [{\citenamefont {Olejn{\'\i}k}\ \emph {et~al.}(2012)\citenamefont
  {Olejn{\'\i}k}, \citenamefont {Wunderlich}, \citenamefont {Irvine},
  \citenamefont {Campion}, \citenamefont {Amin}, \citenamefont {Sinova},\ and\
  \citenamefont {Jungwirth}}]{Olejnik:2012gj}%
  \BibitemOpen
  \bibfield  {author} {\bibinfo {author} {\bibfnamefont {K.}~\bibnamefont
  {Olejn{\'\i}k}}, \bibinfo {author} {\bibfnamefont {J.}~\bibnamefont
  {Wunderlich}}, \bibinfo {author} {\bibfnamefont {A.~C.}\ \bibnamefont
  {Irvine}}, \bibinfo {author} {\bibfnamefont {R.~P.}\ \bibnamefont {Campion}},
  \bibinfo {author} {\bibfnamefont {V.~P.}\ \bibnamefont {Amin}}, \bibinfo
  {author} {\bibfnamefont {J.}~\bibnamefont {Sinova}}, \ and\ \bibinfo {author}
  {\bibfnamefont {T.}~\bibnamefont {Jungwirth}},\ }\href@noop {} {\bibfield
  {journal} {\bibinfo  {journal} {Phys. Rev. Lett.}\ }\textbf {\bibinfo
  {volume} {109}},\ \bibinfo {pages} {076601} (\bibinfo {year}
  {2012})}\BibitemShut {NoStop}%
\bibitem [{\citenamefont {Yu}\ and\ \citenamefont
  {Flatt{\'e}}(2002)}]{Yu:2002gm}%
  \BibitemOpen
  \bibfield  {author} {\bibinfo {author} {\bibfnamefont {Z.~G.}\ \bibnamefont
  {Yu}}\ and\ \bibinfo {author} {\bibfnamefont {M.~E.}\ \bibnamefont
  {Flatt{\'e}}},\ }\href@noop {} {\bibfield  {journal} {\bibinfo  {journal}
  {Phys. Rev. B}\ }\textbf {\bibinfo {volume} {66}},\ \bibinfo {pages} {235302}
  (\bibinfo {year} {2002})}\BibitemShut {NoStop}%
\bibitem [{\citenamefont {Vera-Marun}\ \emph {et~al.}(2011)\citenamefont
  {Vera-Marun}, \citenamefont {Ranjan},\ and\ \citenamefont {van
  Wees}}]{VeraMarun:2011bn}%
  \BibitemOpen
  \bibfield  {author} {\bibinfo {author} {\bibfnamefont {I.~J.}\ \bibnamefont
  {Vera-Marun}}, \bibinfo {author} {\bibfnamefont {V.}~\bibnamefont {Ranjan}},
  \ and\ \bibinfo {author} {\bibfnamefont {B.~J.}\ \bibnamefont {van Wees}},\
  }\href@noop {} {\bibfield  {journal} {\bibinfo  {journal} {Phys. Rev. B}\
  }\textbf {\bibinfo {volume} {84}},\ \bibinfo {pages} {241408} (\bibinfo
  {year} {2011})}\BibitemShut {NoStop}%
\bibitem [{\citenamefont {Vera-Marun}\ \emph {et~al.}(2012)\citenamefont
  {Vera-Marun}, \citenamefont {Ranjan},\ and\ \citenamefont {van
  Wees}}]{VeraMarun:2012kk}%
  \BibitemOpen
  \bibfield  {author} {\bibinfo {author} {\bibfnamefont {I.~J.}\ \bibnamefont
  {Vera-Marun}}, \bibinfo {author} {\bibfnamefont {V.}~\bibnamefont {Ranjan}},
  \ and\ \bibinfo {author} {\bibfnamefont {B.~J.}\ \bibnamefont {van Wees}},\
  }\href@noop {} {\bibfield  {journal} {\bibinfo  {journal} {Nature Phys.}\
  }\textbf {\bibinfo {volume} {8}},\ \bibinfo {pages} {313} (\bibinfo {year}
  {2012})}\BibitemShut {NoStop}%
\bibitem [{\citenamefont {Paget}(1981)}]{Paget:1981td}%
  \BibitemOpen
  \bibfield  {author} {\bibinfo {author} {\bibfnamefont {D.}~\bibnamefont
  {Paget}},\ }\href@noop {} {\bibfield  {journal} {\bibinfo  {journal} {Phys.
  Rev. B}\ }\textbf {\bibinfo {volume} {24}},\ \bibinfo {pages} {3776}
  (\bibinfo {year} {1981})}\BibitemShut {NoStop}%
\bibitem [{\citenamefont {K{\"o}lbl}\ \emph {et~al.}(2012)\citenamefont
  {K{\"o}lbl}, \citenamefont {Zumb{\"u}hl}, \citenamefont {Fuhrer},
  \citenamefont {Salis},\ and\ \citenamefont {Alvarado}}]{Kolbl:2012gi}%
  \BibitemOpen
  \bibfield  {author} {\bibinfo {author} {\bibfnamefont {D.}~\bibnamefont
  {K{\"o}lbl}}, \bibinfo {author} {\bibfnamefont {D.~M.}\ \bibnamefont
  {Zumb{\"u}hl}}, \bibinfo {author} {\bibfnamefont {A.}~\bibnamefont {Fuhrer}},
  \bibinfo {author} {\bibfnamefont {G.}~\bibnamefont {Salis}}, \ and\ \bibinfo
  {author} {\bibfnamefont {S.~F.}\ \bibnamefont {Alvarado}},\ }\href@noop {}
  {\bibfield  {journal} {\bibinfo  {journal} {Phys. Rev. Lett.}\ } (\bibinfo
  {year} {2012})}\BibitemShut {NoStop}%
\bibitem [{\citenamefont {Chan}\ \emph {et~al.}(2009)\citenamefont {Chan},
  \citenamefont {Hu}, \citenamefont {Zhang}, \citenamefont {Kondo},
  \citenamefont {Palmstr{\o}m},\ and\ \citenamefont {Crowell}}]{Chan:2009bu}%
  \BibitemOpen
  \bibfield  {author} {\bibinfo {author} {\bibfnamefont {M.~K.}\ \bibnamefont
  {Chan}}, \bibinfo {author} {\bibfnamefont {Q.~O.}\ \bibnamefont {Hu}},
  \bibinfo {author} {\bibfnamefont {J.}~\bibnamefont {Zhang}}, \bibinfo
  {author} {\bibfnamefont {T.}~\bibnamefont {Kondo}}, \bibinfo {author}
  {\bibfnamefont {C.~J.}\ \bibnamefont {Palmstr{\o}m}}, \ and\ \bibinfo
  {author} {\bibfnamefont {P.~A.}\ \bibnamefont {Crowell}},\ }\href@noop {}
  {\bibfield  {journal} {\bibinfo  {journal} {Phys. Rev. B}\ }\textbf {\bibinfo
  {volume} {80}},\ \bibinfo {pages} {161206} (\bibinfo {year}
  {2009})}\BibitemShut {NoStop}%
\bibitem [{\citenamefont {Paget}\ \emph {et~al.}(1977)\citenamefont {Paget},
  \citenamefont {Lampel}, \citenamefont {Sapoval},\ and\ \citenamefont
  {Safarov}}]{Paget:1977ge}%
  \BibitemOpen
  \bibfield  {author} {\bibinfo {author} {\bibfnamefont {D.}~\bibnamefont
  {Paget}}, \bibinfo {author} {\bibfnamefont {G.}~\bibnamefont {Lampel}},
  \bibinfo {author} {\bibfnamefont {B.}~\bibnamefont {Sapoval}}, \ and\
  \bibinfo {author} {\bibfnamefont {V.~I.}\ \bibnamefont {Safarov}},\
  }\href@noop {} {\bibfield  {journal} {\bibinfo  {journal} {Phys. Rev. B}\
  }\textbf {\bibinfo {volume} {15}},\ \bibinfo {pages} {5780} (\bibinfo {year}
  {1977})}\BibitemShut {NoStop}%
\bibitem [{\citenamefont {Crooker}\ \emph {et~al.}(2009)\citenamefont
  {Crooker}, \citenamefont {Garlid}, \citenamefont {Chantis}, \citenamefont
  {Smith}, \citenamefont {Reddy}, \citenamefont {Hu}, \citenamefont {Kondo},
  \citenamefont {Palmstr{\o}m},\ and\ \citenamefont
  {Crowell}}]{Crooker:2009ju}%
  \BibitemOpen
  \bibfield  {author} {\bibinfo {author} {\bibfnamefont {S.~A.}\ \bibnamefont
  {Crooker}}, \bibinfo {author} {\bibfnamefont {E.~S.}\ \bibnamefont {Garlid}},
  \bibinfo {author} {\bibfnamefont {A.~N.}\ \bibnamefont {Chantis}}, \bibinfo
  {author} {\bibfnamefont {D.~L.}\ \bibnamefont {Smith}}, \bibinfo {author}
  {\bibfnamefont {K.~S.~M.}\ \bibnamefont {Reddy}}, \bibinfo {author}
  {\bibfnamefont {Q.~O.}\ \bibnamefont {Hu}}, \bibinfo {author} {\bibfnamefont
  {T.}~\bibnamefont {Kondo}}, \bibinfo {author} {\bibfnamefont {C.~J.}\
  \bibnamefont {Palmstr{\o}m}}, \ and\ \bibinfo {author} {\bibfnamefont
  {P.~A.}\ \bibnamefont {Crowell}},\ }\href@noop {} {\bibfield  {journal}
  {\bibinfo  {journal} {Phys. Rev. B}\ }\textbf {\bibinfo {volume} {80}},\
  \bibinfo {pages} {041305} (\bibinfo {year} {2009})}\BibitemShut {NoStop}%
\bibitem [{\citenamefont {Ashcroft}\ and\ \citenamefont
  {Mermin}(1976)}]{Ashcroft:1976ud}%
  \BibitemOpen
  \bibfield  {author} {\bibinfo {author} {\bibfnamefont {N.~W.}\ \bibnamefont
  {Ashcroft}}\ and\ \bibinfo {author} {\bibfnamefont {N.~D.}\ \bibnamefont
  {Mermin}},\ }\href@noop {} {\emph {\bibinfo {title} {{Solid State
  Physics}}}}\ (\bibinfo  {publisher} {Holt, Rinehart and Winston},\ \bibinfo
  {address} {New York},\ \bibinfo {year} {1976})\BibitemShut {NoStop}%
\bibitem [{\citenamefont {Johnson}\ and\ \citenamefont
  {Lark-Horovitz}(1953)}]{Johnson:1953ud}%
  \BibitemOpen
  \bibfield  {author} {\bibinfo {author} {\bibfnamefont {V.~A.}\ \bibnamefont
  {Johnson}}\ and\ \bibinfo {author} {\bibfnamefont {K.}~\bibnamefont
  {Lark-Horovitz}},\ }\href@noop {} {\bibfield  {journal} {\bibinfo  {journal}
  {Physical Review}\ }\textbf {\bibinfo {volume} {92}},\ \bibinfo {pages} {226}
  (\bibinfo {year} {1953})}\BibitemShut {NoStop}%
\bibitem [{\citenamefont {Smith}(1978)}]{Smith:1978wp}%
  \BibitemOpen
  \bibfield  {author} {\bibinfo {author} {\bibfnamefont {R.~A.}\ \bibnamefont
  {Smith}},\ }\href@noop {} {\emph {\bibinfo {title} {{Semiconductors}}}},\
  \bibinfo {edition} {2nd}\ ed.\ (\bibinfo  {publisher} {Cambridge University
  Press},\ \bibinfo {address} {New York},\ \bibinfo {year} {1978})\BibitemShut
  {NoStop}%
\bibitem [{\citenamefont {Isenberg}\ \emph {et~al.}(1948)\citenamefont
  {Isenberg}, \citenamefont {Russell},\ and\ \citenamefont
  {Greene}}]{Isenberg:1948jk}%
  \BibitemOpen
  \bibfield  {author} {\bibinfo {author} {\bibfnamefont {I.}~\bibnamefont
  {Isenberg}}, \bibinfo {author} {\bibfnamefont {B.~R.}\ \bibnamefont
  {Russell}}, \ and\ \bibinfo {author} {\bibfnamefont {R.~F.}\ \bibnamefont
  {Greene}},\ }\href@noop {} {\bibfield  {journal} {\bibinfo  {journal} {Rev.
  Sci. Instrum.}\ }\textbf {\bibinfo {volume} {19}},\ \bibinfo {pages} {685}
  (\bibinfo {year} {1948})}\BibitemShut {NoStop}%
\bibitem [{Note1()}]{Note1}%
  \BibitemOpen
  \bibinfo {note} {See Supplemental Material at {[}URL from Publisher{]} for
  calculation details.}\BibitemShut {Stop}%
\end{thebibliography}%

\end{document}